\documentstyle{mn}
\input{psfig}

\def\littleprime{\ifmmode{\scriptscriptstyle \prime }
    \else{\hbox{$\scriptscriptstyle \prime$ }}\fi}
\def\littlecirc{\ifmmode{\scriptscriptstyle \circ }
    \else{\hbox{$\scriptscriptstyle \circ $ }}\fi}
\def\littless{\ifmmode{\scriptscriptstyle s }
    \else{\hbox{$\scriptscriptstyle s $ }}\fi}
\def\arcsec{\raise .9ex \hbox{\littleprime\hskip-3pt\littleprime}}
\def\arcmin{\raise .9ex \hbox{\littleprime}}
\def\degree{\raise .9ex \hbox{\littlecirc}}
\def\arcsecpoint{\hbox to 1pt{}\rlap{\arcsec}.\hbox to 2pt{}}
\def\arcminpoint{\hbox to 1pt{}\rlap{\arcmin}.\hbox to 2pt{}}
\def\degreepoint{\hbox to 1pt{}\rlap{\degree}.\hbox to 2pt{}}
\def\gtapr {\lower .1ex\hbox{\rlap{\raise .6ex\hbox{\hskip .3ex
        {\ifmmode{\scriptscriptstyle >}\else
                {$\scriptscriptstyle >$}\fi}}}
        \kern -.4ex{\ifmmode{\scriptscriptstyle \sim}\else
                {$\scriptscriptstyle\sim$}\fi}}}
\def\ltapr {\lower .1ex\hbox{\rlap{\raise .6ex\hbox{\hskip .3ex
        {\ifmmode{\scriptscriptstyle <}\else    
                {$\scriptscriptstyle <$}\fi}}}
        \kern -.4ex{\ifmmode{\scriptscriptstyle \sim}\else
                {$\scriptscriptstyle\sim$}\fi}}}
\title{IRAM observations of JVAS/CLASS gravitational lenses}
\author[E. Xanthopoulos et al.] {E.~Xanthopoulos$^{1}$, F.~Combes$^{2}$, T.~Wiklind$^{3}$
\\
$^{1}$University of Manchester, Jodrell Bank Observatory,
Macclesfield, Cheshire SK11 9DL, England\\ 
$^{2}$ DEMIRM, Observatoire de Paris, 61 Av. de l'Observatoire, F-75014 Paris, France \\
$^{3}$ Onsala Space Observatory, S-43992 Onsala, Sweden}

\date{Accepted  . 
     Received: }

\begin{document}
\maketitle

\begin{abstract}
We have searched for molecular absorption lines at millimeter wavelengths in eleven  
gravitational lens systems discovered in the JVAS/CLASS surveys of flat spectrum radio
sources. Spectra of only one source 1030+074 were obtained in the 3-, 2- and 1.3-millimeter 
band at the frequencies corresponding to common molecular transitions of CO and HCO$^{+}$
as continuum emission was not found in any of the other sources.
We calculated upper limits to the column density in molecular absorption for 1030+074,
using an excitation temperature of 15 K, to be {\it N$_{\rm CO}$} $<$ 6.3 $\times$ 10$^{13}$ cm$^{-2}$
and {\it N$_{\rm HCO^{+}}$} $<$ 1.3 $\times$ 10$^{11}$ cm$^{-2}$, equivalent to hydrogen column density 
of the order {\it N$_{\rm H}$} $<$ 10$^{18}$ cm$^{-2}$, assuming standard molecular abundances.
We also present the best upper limits of the continuum at the lower frequency for the 
other 10 gravitational lenses. 
  
\end{abstract}

\begin{keywords}
cosmology: observations -- gravitational lensing.
\end{keywords}

\footnotetext{Contact e-mail: emily@jb.man.ac.uk}

\section{Introduction}
Molecular gas, the cold and dense part of the interstellar medium
(ISM) is directly related to the formation of stars. When studied at high
redshifts, it can give us some insight into the evolution of star formation conditions. 
The problem of galaxy evolution is directly related to the study of star 
forming activity and gas content of distant galaxies. However, given the
present-day size and technology of millimeter and centimeter radio telescopes,
even very luminous galaxies are exceedingly difficult to detect at large 
distances. The current sensitivity of radio telescopes is at least 
an order of magnitude below what would be needed to determine the molecular
content of high redshift spirals. This order of magnitude is what is 
gained by observing gravitational lensed objects with single dish radio telescopes 
since magnification ``boosts" the flux of the sources that would otherwise be 
too faint to be studied.

CO emission has been detected in a number of
gravitationally lensed systems (Brown \& Vanden Bout 1991, Barvainis
et al. 1994, Casoli et al. 1996, Alloin et al. 1997, Barvainis et al. 1998).
In 1997, Scoville et al. reported the detection of the first non-lensed 
object at z=2.394 and Guilloteau et al. (1997) of the first radio-quiet quasar  
at z=4.407 with no direct indication of lensing.  
More recently, Papadopoulos et al. (2000) reported the detection of submillimeter 
emission from dust at 850 $\mu$m and of the CO(4-3) line in two high redshift
powerful radio galaxies (see also Combes (1999) for a review of molecular studies 
of high redshift objects). 

\begin{table*}
\begin{center}
\caption{General characteristics of the JVAS/CLASS gravitational lenses in the sample}
\begin{tabular}{llllllllc}\hline
Source & No. of & Source & Lens & Max.  &
\multicolumn{3}{c}{Magnitude} & Refer.\\
            & images & redshift z$_{\rm s}$  & redshift z$_{\rm l}$
&separ. &V & I & H  & \\
\hline
{\sc class}0712+472  & 4&  1.34  & 0.406 & 1.27   &22.4 & 20.0 & 17.7 & (1) \\
{\sc class}0827+525?  & 2& (2.064)    &   & 2.85   &&&& (2) \\
{\sc jvas}1030+074  &   2& 1.53   & 0.599 & 1.56   &22? &20.4 && (3) \\
{\sc class}1127+385  &     2&    &   & 0.701   &24.4& 22.5 && (4) \\
{\sc jvas}1422+231  &   4&  3.62  & 0.34  & 1.28   &21.4 &&& (5) \\
{\sc class}1600+434  &   2&  1.57  & 0.415  & 1.39   &23? &21.2 &18.5& (6)\\
{\sc class}1608+656  &  4&  1.39  & 0.64  & 2.08   &21.4 &19.0 & & (7) \\
{\sc class}1933+503  & 4+4+2 &  (2.62)     & 0.755 & 1.17   &$<$23.5 &21.8 &18.6
& (8) \\
{\sc jvas}1938+666  &  4+2 &       & (0.878)    & 0.93   &    &  &18.5 & (9) \\
{\sc class}2045+265  &  4& 1.28  &  0.87   & 1.86    &    &  &20.4 & (10) \\
{\sc jvas}2114+022  & 2 &       & 0.315   & 2.57  &    & &17.0,17.2 & (11) \\
\hline
\end{tabular}
\label{chara}
\end{center}
\begin{flushleft}
(1) Jackson et al. (1998) \\
(2) Koopmans et al. (2000) \\
(3) Xanthopoulos et al. (1998) \\
(4) Koopmans et al. (1999) \\
(5) Patnaik et al. (1992a) \\
(6) Jackson et al. (1995) \\
(7) Myers et al. (1995) \\
(8) Sykes et al. (1998) \\
(9) Patnaik et al (1992b); King et al. (1997) \\
(10) Fassnacht et al. (1999) \\
(11) King et al. (1999);  Augusto et al. (2000, in preparation)
\end{flushleft}
\end{table*}

\begin{table*}
\caption{Non-detection of absorption in 1030+074 $-$ Upper limits at 3$\sigma$}
\begin{tabular}{rlccccc}\hline
Transition &  $\nu_{\rm obs}$  & {\it T}$_{\rm cont}$ & $\delta\nu$ & $\sigma_{\rm rms}$ &
$\int\tau_{\nu}d\nu$ & {\it N}$_{\rm tot}$\\
       &    GHz    & mK & km s$^{-1}$ & mK & km s$^{-1}$ & cm$^{-2}$ \\
 &   &   &  &  &   &  \\ \hline
HCO+(2-1) & 111.5  & 41  & 2.7 & 2 & 0.4  & $<$ 1.3 $\times$ 10$^{11}$ \\
CO(2-1) & 144.2  & 34  & 2.7 & 2 & 0.5  & $<$ 6.3 $\times$ 10$^{13}$  \\
CO(3-2) & 216.3  & 15  & 2.7 & 3 & 2.4  &  \\ \hline
\end{tabular}
\label{res}
\end{table*}

Absorption molecular lines are much easier to detect, especially at
large distances, since for their detection the sensitivity 
depends on the strength of the background continuum source
and not on the distance. This has been demonstrated by the detection
of more than 40 different molecular transitions at redshifts ranging
from z=0.25 to z=0.89 (Wiklind \& Combes, 1994, 1996a, 1996b; Combes
\& Wiklind 1995). Just as in the optical wavelength band, it is easier
to detect molecular absorption lines (at mm wavelengths) than the
corresponding emission lines. With an appropriate alignment of an
intervening galaxy and a sufficiently strong background radio
continuum source, it is possible to probe the molecular ISM at very
large distances. 
Absorption of molecular rotational lines in the millimeter band is 
then a very sensitive method of studying the physical and chemical 
conditions of molecular gas at large distances.
Also due to the greater sensitivity of the absorption
lines, it is possible to observe molecules other than CO.
These types of surveys can then provide us with valuable information 
about the classification of lenses (see for example Wiklind and Combes (1995)
and their study of the gravitational lens system B0218+357).

The probability of detecting molecular absorption lines is larger when
the impact parameter between the background source and the intervening
galaxy is very small, as is the case in gravitational lens systems. In
fact, Wiklind and Combes (1995, 1996) have shown that it is feasible
to measure the redshift of the lensing object by mm-wave
spectroscopy. This method is extremely promising to provide the much
needed redshifts that will exploit the potential of gravitational
lenses as powerful constraints on cosmology and galactic structure (e.g. Kochanek 1996; 
Keeton et al. 1998; Falco et al. 1998; Helbig, et al. 1999). 
Since we also have multiple lines of sight, this type of surveys gives us
the possibility of useful kinematical information.

\begin{table}
\begin{center}
\caption{1030+074: The centimeter and millimeter fluxes}
\begin{tabular}{ll}\hline
 Frequency  & Flux  \\
 (GHz)    & (mJy)    \\
         &     \\ \hline
0.365    &  199 $\pm$ 9.0   \\
1.4    &   155 $\pm$ 4.0   \\
4.85  &    341 $\pm$ 5.0  \\
8.4 &     203 $\pm$ 0.3 \\
111.5 &   246 $\pm$  12  \\
144.2 &   255 $\pm$  12  \\
216.3   &  144 $\pm$  10    \\ \hline
\end{tabular}
\label{fluxall}
\end{center}
\end{table}

The uses of molecular absorption lines are manifold. Apart from studying the physical and 
chemical properties of molecular gas in distant galaxies, they can be used as probes of 
very small scale structures in the molecular gas. They can be used as cosmographic probes,
to set upper limits to the temperature of the cosmic microwave background radiation and to 
constrain the geometry in gravitationally lensed systems.  

The Jodrell-Bank VLA Astrometric Survey (JVAS; Patnaik et al. 1992a; Patnaik 1993; Browne et al. 1998, 
Wilkinson et al. 1998; King et al. 1999) 
and the Cosmic Lens All Sky Survey (CLASS; Browne et al. 1999; 
Myers et al. 1999)
are surveys of flat-spectrum radio sources one of whose purposes is to search for
gravitational lens systems. Nineteen new lenses have been uncovered up to now. 
In this paper we present the results of submillimeter observations of 11 of these lens systems. 
We did not detect absorption in any of the sources and continuum was found in only the 
lens system 1030+074. In Section 2 we describe the observations and in Section 3 we derive 
upper limits to the molecular absorption column densities in 1030+074 and upper limits to 
the continuum emission of the rest of the sources. A discussion follows in Section 4. 

\section{Observations}

The observations were made with the Institut de Radio Astronomie Millim\'{e}trique (IRAM) 
30-m telescope at Pico Veleta, Granada, Spain in 1997 December. Table~\ref{chara} shows the 
main characteristics of the lenses in the sample. In column 1 we give the JVAS/CLASS name of the 
source and in columns 2, 3, 4 and 5 we present the number of components in each 
system, the source and lens redshift, when available (we have put in a parenthesis the redshifts 
that were unknown at the time of the observations), and the maximum separation between 
the components of each system respectively. In columns 6, 7 and 8, one can find the HST/WFPC2 
V and I and HST/NICMOS H magnitudes (Impey et al. 1996; King et al. 1998; Marlow et al. 1999; 
Jackson et al. 2000; and references in Table~\ref{chara}) 
of the lensing galaxy. The discovery paper reference of 
each system is given in the last column of the table.  
We used the 3, 2 and 1.3 mm SIS receivers.
The observations were done with a nutating subreflector with a frequency of 0.5 Hz and with a 
throw of $\pm$90\arcsec in azimuth. We used a frequency resolution of 1 MHz, which 
corresponds to a velocity resolution of 2.7 km s$^{-1}$ at 3-mm.  
The continuum levels of the observed sources were 
determined using a continuum backend and increasing the subreflector switch frequency to 
2 Hz. Since none of the 11 systems/sources had a known continuum millimeter flux we decided
to do the following: for the three lenses in the sample for which we already had a confirmed 
redshift and were known to have a strong continuum at 4.85 GHz (namely 1030+074, 1422+231, and 
2114+022) we tuned the receivers to the redshifted molecular absorption lines of CO and 
HCO$^{+}$. For the other eight sources, for which we did not have any redshift information 
we decided to concentrate initially on the observations of the continuum emission; for 
seven of those the receiver was tuned to the redshift of 1030+074, while for 0712+472
(which was observed in a different programme) the receiver was tuned to the redshift of
0218+357. 

\begin{figure*}
\begin{center}
\setlength{\unitlength}{1cm}
\begin{picture}(5,10)
\put(-1.3,0){\includegraphics{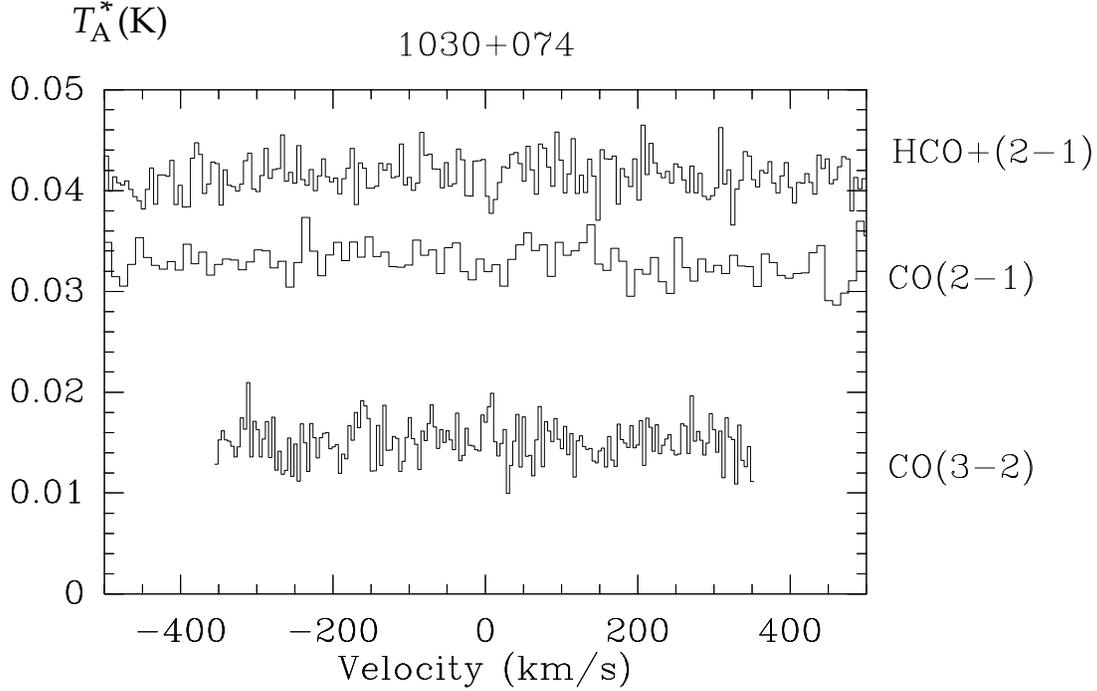}}
\end{picture}
\caption{$\Delta\upsilon$ = 2.7 km s$^{-1}$ spectra of CO (J = 2$-$1), CO (J = 3$-$2) and
HCO$^{+}$(J = 2$-$1) at 1030+074. The velocity scale is heliocentric and centered at
z$=$0.599.}
\label{1030map}
\end{center}
\end{figure*}

The main purpose of this exercise was to 
get knowledge of the continuum level of all sources; 
and if the continuum was large enough (at least 30 mK) to undertake a search of absorption lines,
if necessary by sweeping up the receiver frequency range, 
when the redshift was not accurately known.
We detected continuum emission in 
only one source 1030+074 (see Fig.~\ref{1030map}), but did not find absorption in its spectrum. 
Details of the results are given in the following section. 

\section{Results $-$ Analysis of absorption limits}

\subsection{Molecular absorption upper limits}

We have used the non-detection of absorption in 1030+074 to place upper limits on the 
column density in the various molecules. The calculations were made in the same way as 
in Wiklind \& Combes (1996b), based on the noise in a single spectral element,
1 MHz channels (2.7 km sec$^{-1}$ channels), and the results for 1030+074 are shown 
in Table~\ref{res}. We have assumed an excitation temperature of 15 K. 
In column 1 we have the molecular line, in column 2 the observed frequency and 
in column 3 the measured continuum temperature. In columns 4 and 5 one can find 
the spectral element and the the noise level for each observation. In column 6 we
calculate an integrated optical depth $\int\tau d\nu$ using the formula 
$\tau_{\nu}$$=$$-$ln(1-3$\sigma_{\rm rms}$/{\it T}$_{\rm cont}$). Finally, the upper limits to the 
column densities for the two molecular species are presented in the last column. 
The derived 
upper limit of the absorption is {\it N}$_{\rm CO}$ $<$ 6.3 $\times$ 10$^{13}$ cm$^{-2}$. 
This value is much lower than the typical value of 10$^{15}$$-$10$^{16}$ cm$^{-2}$ 
for molecular gas in a variety of different environments, 
both in our own and nearby galaxies. For the {\it N}$_{\rm HCO}$ we find $<$ 1.3 $\times$ 
10$^{11}$ cm$^{-2}$
at 3$\sigma$. 
If we assume 
standard molecular abundances then we find that the hydrogen column density at the 3$\sigma$ level
is {\it N}$_{\rm H}$ $<$ 10$^{18}$ cm$^{-2}$. 

We convert the values found for the continuum emission of 1030+074 to mJy by using the 
conversion factor, the flux density to antenna temperature ratio {\it S/T}$^{*}_{\rm A}$ for 
a point source. These values are 6.0, 7.5 and 9.6 Jy/K for the 3, 2, and 1.3 mm receivers 
respectively. So the continuum emission values that we get for 1030+074 are 
246, 255, and 144 mJy at the three frequencies respectively. In Table~\ref{fluxall} we  
combine our own continuum data with those of lower frequencies available for 1030+074.  
The lower frequency 0.365 GHz, 1.4 GHz 
and 4.85 GHz flux densities were obtained from the Texas (Douglas et al. 1996), NVSS (Condon et al. 1998) 
and GB6 (Gregory et al. 1996) catalogues/surveys
and are 199 mJy, 155 mJy and 341 mJy respectively. The 8.4 GHz flux density of 203.3 mJy was 
obtained from the CLASS VLA data. Table~\ref{fluxall} gives a further indication that 1030+074 
is a system with strong variability (Xanthopoulos et al. 2000, in preparation) and so a 
source suitable for H$_{0}$ determination.    

For the other 10 lens systems for which we did not detect continuum emission, within 15 min of 
observing time on each source, we present in 
Table~\ref{nondetect} the best upper limit in mK and mJy at the lower 3-mm frequency. 

We also present in Table~\ref{allinfo} a collection of all the radio fluxes available for these
systems from the Texas, NVSS, GB6, WENSS (Rengelink et al. 1997) and other published in  
the literature. In the last column we also note whether variability was detected or not 
in these sources. Most of the radio information come from the discovery papers that can be 
found in the last column of Table~\ref{chara}.    

\subsection{Optical absorption}
The lens system 1030+074 has been observed with the HST WFPC2 and NICMOS at V, I and H 
(Jackson et al. 2000). Falco et al. (1999) have determined the total E(B$-$V) extinction for 
this lens system to be 0.38$\pm$0.18 which we use to convert to a hydrogen 
column density.  
For this conversion we use the Bohlin et al. (1978) relation that gives the mean ratio of the total neutral 
hydrogen to the color excess, that is: N(H~I $+$ H$_{2}$)/E(B$-$V) $=$ 5.8 $\times$ 10$^{21}$ 
atoms cm$^{-2}$ mag$^{-1}$ for the Solar neighbourhood.  
This relationship gives \\
\begin{center}{{\it N}$_{\rm H} \approx $ 2.2 $\times$ 10$^{21}$ cm$^{-2}$.}\end{center}
So if all the reddening is dust at the lens redshift, then we have a hydrogen column density 
which is 3 orders of magnitude larger that the upper limit implied by the molecular absorption data.
One possible explanation might be that the total extinction is associated with the source rather
than the lens galaxy.  Another may be that the column density is underestimated by 
three orders of magnitude because the molecular ISM is made up of small dense clumps rather than 
a diffuse medium.  

\begin{table}
\begin{center}
\caption{Upper limit of the continuum at the lower frequency in JVAS/CLASS gravitational lenses. The
upper limits are given at 1 $\sigma$$_{\rm rms}$ in 1 MHz channels.}
\begin{tabular}{crrr}\hline
Source &  Frequency  & {\it T}$_{\rm cont}$  & Flux \\
       &    (GHz)    & (mK) & (mJy) \\
 &   &   &  \\ \hline
0712+472 & 85  & 15 &  90 \\
0827+525 & 111 & 27 & 162 \\
1127+385 & 111 &  8 &  48 \\
1422+231 & 108 &  7 &  42 \\
1600+434 & 111 & 19 & 114 \\
1608+656 & 111 & 11 &  66 \\
1933+503 & 111 & 19 & 114 \\
1938+666 & 111 & 19 & 114 \\
2045+265 & 111 & 19 & 114 \\
2114+022 & 87  &   5  & 30 \\ \hline
\end{tabular}
\label{nondetect}
\end{center}
\end{table}

Jackson et al. (2000) have already suggested that taking into consideration the fact that 
the H band ratio of the component images is larger than the V and I flux density ratios, 
microlensing is a possibility for this system. 
The colours of the lensed images are consistent with our knowledge that the lensed object 
is a quasar/BL Lac type object (Fassnacht \& Cohen 1998). 
The most plausible explanation, however, is that the lens extinction
involves only the B-image, since Falco et al. (1999) only determined the
differential extinction between B and A, and B is the absorbed one. Our
upper limit on H$_{2}$ column density only concerns the A-image, because of the 
large continuum ratio between A and B. If this ratio is the same at
cm and mm wavelengths, we expect the continuum towards B of the order of
2 or 3 mK, i.e. of the order of 1$\sigma$ level of our 3-mm spectrum.
Therefore, we are not sensitive to even a high column density towards B.
The fact that the molecular column density is patchy over a galaxy, and
only one of the images is covered and not the other, is frequent as
already determined in PKS1830-211 and B0218+357.

\section{Discussion}
The non-detection of continuum in 10 of the 11 lens systems that we observed 
shows that they are weaker at mm wavelengths than at radio wavelengths. This is not 
entirely unexpected since the sources were selected to have flat spectral indices 
($\alpha > -$0.5, {\it S}$\propto \nu^{\alpha}$) between 1.4 and 5 GHz. 
As can be seen from Table~\ref{allinfo} however, a lot of the sources in our sample show 
quite a steep decrease in their continuum flux at 8.4 GHz. 
Also, the spectral index contains an uncertainty from the fact that many 
of the selected sources are variable. The only source that actually seems to have a 
strong enough continuum, and despite known variability with frequency up to 20-30\% (Xanthopoulos et al. 1998) 
and signs of variability with time maybe up to 30-40\% (Xanthopoulos et al. in preparation),
to be detected even at the submm wavelengths is 1030+074. 

\begin{table*}
\begin{center}
\caption{Radio information of the lens systems in our sample.}
\begin{tabular}{crrrrrrc}\hline
Source & \multicolumn{6}{c}{Radio fluxes (mJy)} & Variability \\
       &  0.325 GHz & 0.365 GHz & 1.4 GHz & 4.85 GHz & 8.4 GHz & 15 GHz &  \\
 &   &   & & &  &   \\ \hline
0712+472 & 41 & -- & 25 & 30 & 26.6  & -- &  yes \\
0827+525 & 71 & -- & 72 & 47 & 28.0  & 37.1 & yes \\
1127+385 & 12  &    & 29.3    & 15.3   & 14.7      &    & no \\
1422+231 & -- & -- & 268 & 548 & 152.8   & -- & yes \\
1600+434 & 37    &    &  76.4   & 37    & 132       &   &  yes \\
1608+656 & 228 & -- & 67 & 88 & 35.1  &  -- &  yes \\
1933+503 & 310 & -- & 108 & 63 & 17.6  & --  & no \\
1938+666 & 753 & 681 & 576 & 329 & 169.9 & --  & ? \\
2045+265 & -- & -- & 55 & 55 & 18.2  &  --  & ? \\
2114+022 & -- & -- & 137 & 224 & 43.4 & --  & ? \\ \hline
\end{tabular}
\label{allinfo}
\end{center}
\end{table*}

The observations of 1030+074 were tuned at the lens redshift. 
The non-detection of absorption in 1030+074 can 
be due to one of the following reasons: 
\begin{itemize}
\item The absorption may not take place at the lens redshift, but in intervening systems at lower 
redshift.  
\item The reddening may not be real, but the result of continuum 
emission that is intrinsically red (e.g. Falco et al. 1997). 
\item Differences in absorption between the components of the lensed image. 
We note that 
1030+074 is the lens system with the largest flux density ratio between its components 
known presently. The flux density ratio ranges between 12 and 19 at centimeter wavelengths.
The fainter B component in the system is 
so close to the lensing galaxy that it is heavily obscured. Component B is fainter by approximately  
3.5-4 mag compared to component A (Xanthopoulos et al. 1998); the optical 
flux ratio of the two components is much greater than the ratio in the radio which is consistent
with extinction of B due to dust from the lensing galaxy (profile fitting to the lensing galaxy 
supports its spiral morphology). 
\end{itemize}

Falco et al. (1999) use the absorption by H~I and molecular gas in the lens galaxy in B0218+357 and 
PKS1830-211 to get direct estimates of the extinction in order to determine the dust-to-gas ratios in the 
lens galaxies in their sample in which 1030+074 is included. From this they note that the molecular
absorption is inferred to be in optically thick CO clouds that incompletely cover the continuum 
radio source. 

In summary, we find no evidence for molecular absorption from the lens system 1030+074
from which we deduce upper limits for the CO and HCO$^{+}$ molecular absorption species.   
We detect continuum emission in only 1 of the 11 lens systems in our sample, 1030+074, in all 3 frequencies.

\section*{Acknowledgments}
We would like to thank our referee Lindsay King for many useful comments.
This research was supported by European Commission, TMR Programme,
Research Network Contract ERBFMRXCT96-0034 ``CERES".

\end{document}